\documentclass[prd,aps,twocolumn,nofootinbib]{revtex4}

\usepackage{graphicx}
\usepackage{epsfig}

\begin{document}

\title{Self-Consistency Requirements of the Renormalization Group for Setting the Renormalization Scale}

\author{Stanley J. Brodsky}
\email{sjbth@slac.stanford.edu}
\affiliation{SLAC National Accelerator Laboratory, Stanford University, Stanford, California 94039, USA}

\author{Xing-Gang Wu}
\email{wuxg@cqu.edu.cn}
\affiliation{Department of Physics, Chongqing University, Chongqing 401331, P.R. China}
\affiliation{SLAC National Accelerator Laboratory, Stanford University, Stanford, California 94039, USA}

\date{\today}

\begin{abstract}

In conventional treatments, predictions from fixed-order perturbative QCD calculations cannot be fixed with certainty due to ambiguities in the choice of the renormalization scale as well as the renormalization scheme. In this paper we present a general discussion of the constraints of the renormalization group (RG) invariance on the choice of the renormalization scale. We adopt the RG based equations, which incorporate the scheme parameters, for a general exposition of RG invariance, since they simultaneously express the invariance of physical observables under both the variation of the renormalization scale and the renormalization scheme parameters. We then discuss the self-consistency requirements of the RG, such as reflexivity, symmetry, and transitivity, which must be satisfied by the scale-setting method. The Principle of Minimal Sensitivity (PMS) requires the slope of the approximant of an observable to vanish at the renormalization point. This criterion provides a scheme-independent estimation, but it violates the symmetry and transitivity properties of the RG and does not reproduce the Gell-Mann-Low scale for QED observables. The Principle of Maximum Conformality (PMC) satisfies all of the deductions of the RG invariance - reflectivity, symmetry, and transitivity. Using the PMC, all non-conformal $\{\beta^{\cal R}_i\}$-terms (${\cal R}$ stands for an arbitrary renormalization scheme) in the perturbative expansion series are summed into the running coupling, and one obtains a unique, scale-fixed, scheme-independent prediction at any finite order. The PMC scales and the resulting finite-order PMC predictions are both to high accuracy independent of the choice of initial renormalization scale, consistent with RG invariance. Moreover, after PMC scale-setting, the residual initial scale-dependence at fixed order due to unknown higher-order $\{\beta_i\}$-terms can be substantially suppressed. The PMC thus eliminates a serious systematic scale error in pQCD predictions, greatly improving the precision of tests of the Standard Model and the sensitivity to new physics at collider and other experiments. \\

\begin{description}

\item[PACS numbers] 12.38.Aw, 12.38.Bx, 11.10.Gh, 11.10.Hi

\end{description}

\end{abstract}

\maketitle


\section{introduction}

Given the perturbative series for a physical quantity
\begin{equation}\label{phyvalue}
\label{sec:intro}
\rho_n = {\cal C}_0 \; \alpha_s^p(\mu) + \sum_{i=1}^{n}{\cal C}_i(\mu) \; \alpha_s^{p+i}(\mu)
\end{equation}
expanded to $n$-th order in the QCD strong coupling constant $\alpha_s(\mu)$, the renormalization scale $\mu$ must be specified in order to obtain a definite prediction. The common practice adopted in the literature is to simply guess a renormalization scale $\mu = Q$, keep it fixed during the calculation ($Q$ is usually assumed to be a typical momentum transfer of the process), and then vary it over an arbitrary range, e.g. $[Q/2, 2\,Q]$, in order to ascertain the scale-uncertainty. However there are many weak points of this conventional scale-setting method :

\begin{enumerate}
\item Although the infinite perturbative series $\rho_{n\to\infty}$ summed to all orders is renormalization-scale independent, the scale dependence from $\alpha_s(\mu)$ and ${\cal C}_i(\mu)$ do not exactly cancel at finite order, leading to a renormalization scale ambiguity.

\item The fixed-order estimate in the conventional procedure is also scheme dependent; i.e., different choice of renormalization scheme ${\cal R}$ will lead to different theoretical estimates. This is the well-known renormalization scheme ambiguity~\cite{ambi1,ambi2,ambi3,FAC1,FAC2,FAC3,PMS1,PMS2,PMS3,PMS4,BLM}.

\item The conventional scale choice can give unphysical results: For example, for the case of $W$-boson plus three-jet production at the hadronic colliders, taking $\mu$ to be the $W$-boson transverse energy, the conventional scale-setting method even predicts negative QCD cross-section at the next-to-leading-order (NLO)~\cite{wjet1,wjet2}.

\item As has been shown in Ref.~\cite{pmc4}, taking an incorrect renormalization scale underestimates the top quark forward-backward asymmetry at the Tevatron.

\item It should be recalled that there is no ambiguity in setting the renormalization scale in QED.
    \begin{itemize}
    \item In QED, the coupling $\alpha(q^2)$ is conventionally defined in the Gell Mann-Low (GM-L) scheme~\cite{gell} from the potential between heavy charges, and it is normalized  at $q^2=0$ to the fine-structure constant $\alpha(0)\simeq 1/137.0359...$~\cite{pdg}.
    \item Due to the Ward-Takahashi identity~\cite{WTidentity}, the divergences in the vertex and fermion wavefunction corrections cancel, and the ultraviolet divergence associated with the vacuum polarization defines a natural scale for the coupling constant $\alpha(q^2)$ with $q^2$ being the squared momentum transfer for the photon propagator. This fact was first observed by Gell-Mann and Low~\cite{gell}; i.e., in the standard GM-L scheme, the renormalization scale is simply the virtuality of the exchanged photon.

        For example, the renormalization scale for the electron-muon elastic scattering based on one-photon exchange is the virtuality of the exchanged photon, i.e. $\mu^2_{\rm GM-L} =t=q^2$. One can of course choose any initial renormalization scale $t_0$ for calculating the QED amplitude; however, the final result will not depend on the choice of $t_0$, since
        \begin{equation}
         \alpha(t) = \frac{\alpha(t_0)}{[1 - \Pi(t,t_0)]} \;, \label{qedsample}
        \end{equation}
        where $$\Pi(t,t_0)= \frac{[\Pi(t) -\Pi(t_0)]}{[1-\Pi(t_0)]}$$ naturally sums all vacuum polarization contributions, both proper and improper, to the dressed photon propagator. (Here $\Pi(t) =\Pi(t,0)$ is the sum of proper vacuum polarization insertions, subtracted at $t=0$.) The invariance of the result on the initial scale $t_0$ is the property used to derive the Callan-Symanzik equations~\cite{callan,symanzik}. There is, therefore, no reason to vary $\mu_{\rm GM-L}$ by a factor of $1/2$ or $2$, since the photon virtuality $t$ is the unique, optimized scale in the GM-L scheme.

    \item The renormalization scale in QED is unique in any scheme including dimensional regularization; different schemes can be connected to the GM-L scheme by commensurate scale relations (CSRs)~\cite{sr}, a topic which we discuss below. The computation of higher-order $\{\beta^{\cal R}_i\}$-functions is thus important for perturbative calculations at the highest orders~\cite{fiveqed1,fiveqed2,fiveqed3}.
    \end{itemize}
    
\item There are uncancelled large logarithms, as well as ``renormalon" terms in higher orders which diverge as $\left[n!(\beta^{\cal R}_i)^{n} \alpha_s^n \right]$~\cite{renormalon}. The convergence of the perturbative series is thus problematic using conventional scale-setting. For certain processes such as the top-quark pair production, it is found that the total cross-section for the $(q\bar{q})$-channel, $q\bar{q}\to t+\bar{t}$, at the next-to-next-to-leading order (NNLO) is about $50\%$ of the NLO cross-section using conventional scale-setting~\cite{pmc3,nnlo1,nnlo2}. Thus, to derive a dependable perturbative estimate, one evidently needs to do even higher order calculations.

\item The conventional estimate shows a strong dependence on the choice of the renormalization scale $\mu$. It is clearly artificial to guess a renormalization scale $\mu = Q$ and to study its uncertainty by simply varying $\mu$ in the arbitrary range $[Q/2, 2\,Q]$. Why is the scale uncertainty estimated only by varying a factor of $1/2$ or $2$, and not, say, $10$ times Q ? For example, Ref.\cite{Q6} shows that after including the first and second order corrections to several deep inelastic sum rules which are due to heavy flavor contributions, it is found that the effective scale $\mu \sim 6.5\, Q$, where the typical scale $Q=m$ with $m$ being the corresponding heavy quark mass.

    Moreover, sometimes, there are several choices for the typical momentum transfer of the process, all of which can be taken as the renormalization scale, such as the heavy-quark mass, the collision energy of the subprocess, etc. Which invariant provides the correct theoretical estimate ?

    Using conventional scale-setting, there is no definite answer to these questions. One may argue that the correct renormalization scale for the fixed-order prediction can be decided by comparing with the experimental data, but this surely is process dependent and greatly depresses the predictive power of the pQCD theory.

\end{enumerate}

Thus, in summary, the conventional scale-setting assigns an arbitrary range and an arbitrary systematic error to fixed-order pQCD predictions. In fact, as we discuss in this article, this {\it ad hoc} assignment of the range and associated systematic error is unnecessary and, in fact, can be eliminated.

\begin{figure}[ht]
\includegraphics[width=0.4\textwidth]{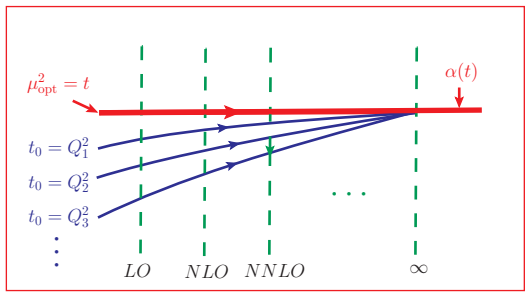}
\caption{Pictorial representation of the optimized renormalization scale $\mu_{\rm opt}$. Taking electron-muon elastic scattering through one-photon exchange as an example: In the GM-L scheme, the optimized scale is $\mu^2_{\rm opt}=t$ which corresponds to the scale-invariant value $\alpha(t)$. As a comparison, the values of $\alpha$ at fixed-orders for different choice of $t_0=Q^2_{i}$ ($i=1,2,3,\cdots$) are shown by thin-and-solid curves. }
\label{optimizalscale}
\end{figure}

One may ask: For a general fixed-order calculation, what is the correct ``physical" scale or optimized scale ? To our understanding, it should provide a prediction independent of the renormalization scheme and the choice of initial scale. A pictorial representation of what is the optimized renormalization scale is shown in Fig.(\ref{optimizalscale}), where the electron-muon elastic scattering through one-photon exchange is taken as an illustration. In the GM-L scheme, the optimized scale $\mu^2_{\rm opt}=t$ which corresponds to the scale-invariant value $\alpha(t)$. Moreover, by using the proper scale-setting method, such as the newly suggested Principle of Maximum conformality (PMC)~\cite{pmc1,pmc2,pmc3,pmc5}, the prediction is also scheme independent and the argument of the coupling in different schemes have the correct displacement. For example, by using the PMC procedure for QED one obtains the correct displacement between the argument of the coupling in the ${\overline{MS}}$ scheme relative to the GM-L scheme at one loop~\cite{MSbar} \footnote{The displacement for higher-order corrections can be obtained by carefully dealing with the differences of the $\{\beta_i\}$-series under different renormalization schemes. }
\begin{equation}\label{displacement}
\alpha_{GM-L}(t)=\alpha_{\overline{MS}}(e^{-5/3}t) \,.
\end{equation}

As a comparison, the values of $\alpha$ at fixed-orders for different choice of $t_0=Q^2_{i}$ ($i=1,2,3,\cdots$) are shown by thin-and-solid curves in Fig.(\ref{optimizalscale}). A particular choice of $t_0$ using conventional scale-setting may lead to a value of $\alpha$ close to $\alpha(t)$, but this would only be a lucky guess and not the correct answer. As one includes higher-and-higher orders, the guessed scale will lead to a better estimate. In fact, when doing the perturbative calculation up to infinite order, any choice of $t_0$ will lead to the correct value $\alpha(t)$ as required by the RG invariance. However, if one chooses $t_0=t$, the complete all-orders result is obtained from the onset.

Does there exist such an optimized renormalization scale for a general process in non-Abelian QCD ? If it does exist, how can one set it at finite order in a systematic and process-independent way ? This is not an easy task. Various scale-setting procedures have been proposed since the 1980's for deriving an optimized scale, such as Fastest Apparent Convergence (FAC)~\cite{FAC1,FAC2,FAC3} \footnote{As argued by Grunberg~\cite{FAC3} and Krasnikov~\cite{FAC4}, it is better to be called as the RG-improved effective coupling method. For simplicity, we retain the name as FAC as suggested by Stevenson~\cite{PMS2}.}, the Principle of Minimum Sensitivity (PMS)~\cite{PMS1,PMS2,PMS3,PMS4}, the Brodsky-Lepage-Mackenzie (BLM)~\cite{BLM} procedure and its extended versions such as the dressed skeleton expansion~\cite{dse1,dse2}, the sequential se-BLM and $x$-BLM methods~\cite{kataev1,kataev2,kataev3}, etc., and the PMC. A short review of FAC, BLM and PMS can be found in Ref.\cite{book}. In principle, the correctness of a scale-setting method can be judged by the experimental data. However, as we shall discuss, there are self-consistency theoretical requirements which shed light on the reliability of the scale-setting method~\cite{self}.

Clearly, the prediction for any physical observable must be independent of the choice of renormalization scheme; this is the central property of the renormalization group (RG) invariance~\cite{callan,symanzik,peter1,peter2,peter3}. As we shall discuss, the RG based equations~\cite{PMS1,PMS2,PMS3,PMS4} which incorporate the scheme parameters provide a convenient way for estimating both the scale- and scheme- dependence of the QCD predictions for a physical process~\cite{PMS1,PMS2,PMS3,PMS4,pmc2,HJLu}. In this paper, we will utilize such RG based equations for a general discussion of the RG-invariance. We will discuss in detail the self-consistency requirements of the RG~\cite{self}, such as reflexivity, symmetry and transitivity, which must be satisfied by a scale-setting method. We will then show whether the scale-setting methods, FAC, BLM/PMC and PMS, satisfy these requirements.

The remaining parts of this paper are organized as follows: in Sec.~II, we give a general demonstration of the RG-invariance with the help of the RG based equations. In Sec.~III, we discuss the self-consistency requirements for a scale-setting method, where a graphical explanation of these requirements is also given. In Sec.~IV and Sec.~V, we present a detailed discussion on PMC and PMS scale-setting methods, respectively. Sec.~VI provides a summary.

\section{Renormalization-group based equations and the renormalization group invariance}
\label{sec:rg}

The scale dependence of the running coupling in gauge theory is controlled by the RG equation
\begin{equation} \label{basic-RG}
\beta^{\cal R}=\frac{\partial}{\partial\ln\mu^2}\left(\frac{\alpha^{\cal R}_s(\mu)}{4\pi}\right) =-\sum_{i=0}^{\infty}\beta^{\cal R}_{i}\left(\frac{\alpha^{\cal R}_s(\mu)}{4\pi}\right)^{i+2} ,
\end{equation}
where the superscript ${\cal R}$ stands for an arbitrary renormalization scheme, such as $MS$ scheme~\cite{MS}, $\overline{MS}$ scheme~\cite{MSbar}, $MOM$ scheme~\cite{MOM}, etc.. Note that the $\beta^{\cal R}_i$-functions for the $MS$ and $\overline{MS}$ schemes are the same~\cite{kataev-beta}. Various terms in $\beta^{\cal R}_0$, $\beta^{\cal R}_1$, $\cdots$, correspond to one-loop and two-loop $\cdots$ contributions respectively. In general, the $\{\beta^{\cal R}_i\}$ are scheme-dependent and depend on the quark mass $m_f$. According to the decoupling theorem, a quark with mass $m_{f}\gg\mu$ can be ignored, and we can often neglect $m_f$-terms when $m_{f}\ll\mu$. Then, for every renormalization scale $\mu$, one can divide the quarks into active ones with $m_f =0$ and inactive ones that can be ignored. Within this framework, it is well-known that the first two coefficients $\beta^{\cal R}_{0,1}$ are universal; i.e., $\beta^{\cal R}_0 \equiv 11-2n_f/3$ and $\beta^{\cal R}_1 \equiv 102-38n_f/3$ for $n_f$-active flavors. Hereafter, we simply write them as $\beta_0$ and $\beta_1$. It is noted that an analytic extension of $\alpha^{\overline{MS}}_s$ which incorporates the finite-mass quark threshold effects into the running of the coupling has been suggested in Ref.~\cite{msext}. However, numerically, it is found that taking finite quark mass effects into account analytically in the running, rather than using a fixed $n_f$ between thresholds, leads to effects of the order of one percent for the one-loop running coupling~\cite{msext}. Here we will work with the conventional $\{\beta^{\cal R}_i\}$-functions.

It will be convenient to use the first two universal coefficients $\beta_0$ and $\beta_1$ to rescale the coupling constant and the scale-parameter in Eq.(\ref{basic-RG}). By rescaling the coupling constant and the scale parameters as~\cite{HJLu}
\begin{displaymath}
a^{\cal R}=\frac{\beta_1}{4\pi\beta_0}\alpha^{\cal R}_s \;\;{\rm and}\;\; \tau_{\cal R}=\frac{\beta^2_0}{\beta_1} \ln\mu^2 ,
\end{displaymath}
one can express the RG equation (\ref{basic-RG}) into a simpler canonical form
\begin{equation}\label{scale0}
\frac{d a^{\cal R}}{d \tau_{\cal R}} = -(a^{\cal R})^2 \left[1+ a^{\cal R} +c^{\cal R}_2 (a^{\cal R})^2+c^{\cal R}_3 (a^{\cal R})^3 +\cdots \right] ,
\end{equation}
where $c^{\cal R}_i = {\beta^{\cal R}_i \beta_0^{i-1}} / {\beta^i_1}$ for $i=2, 3, \cdots$.

As an extension of the ordinary coupling constant, one can define a universal coupling constant $a(\tau,\{c_i\})$ to include the dependence on the scheme parameters $\{c_i\}$, which satisfies the following extended RG based equations \cite{HJLu}
\begin{equation}
\beta(a,\{c_i\}) = \frac{\partial a}{\partial \tau} = -a^2 \left[1+ a +c_2 a^2+c_3 a^3 +\cdots \right] \label{scale}
\end{equation}
\begin{equation}
\beta_n(a,\{c_i\}) = \frac{\partial a}{\partial c_n} = -\beta(a,\{c_i\}) \int_0^{a} \frac{ x^{n+2} dx}{\beta^2(x,\{c_i\})} \, . \label{scheme}
\end{equation}
The scale-equation (\ref{scale}) can be used to evolve the universal coupling function from one scale to another. The scheme-equation (\ref{scheme}), which was first suggested by Stevenson~\cite{PMS2}, can be used to relate the coupling functions under different schemes by changing $\{c_i\}$. A solution of the scale-equation up to four-loop level has been given in Ref.~\cite{pmc2}, which agrees with that of the conventional RG-equation obtained in the literature, cf. Ref.~\cite{fourloop}. By comparing Eq.(\ref{scale0}) with Eq.(\ref{scale}), there exists a value of $\tau=\tau_R$ for which
\begin{equation}\label{asrelation}
a^{\cal R}(\tau_{\cal R})=a(\tau_{\cal R},\{c^{\cal R}_i\}) .
\end{equation}
This shows that any coupling constant $a^{\cal R}(\tau_{\cal R})$ can be expressed by the universal coupling constant $a(\tau,\{c_i\})$ under proper correspondence; i.e. the coupling constant $a^{\cal R}(\tau_{\cal R})$ can be treated as a special case of the universal coupling constant $a(\tau,\{c_i\})$: Any usual coupling constant $a^{\cal R}(\tau_{\cal R})$ is equal to a universal coupling $a(\tau_{\cal R}, \{c_i\})$ by setting $\{c_i\}$ to be $\{c^{\cal R}_i\}$, since both coupling constants satisfy the same RG equation by using the same scheme parameters.

Grunberg has pointed out that~\cite{FAC1,FAC2,FAC3} any perturbatively calculable physical quantity can be used to define an effective coupling constant by incorporating the entire radiative corrections into its definition. The effective coupling constant satisfies the same RG equation (and hence the same RG based equations) as the usual (universal) coupling constant. Thus, the running behavior for both the effective coupling constant and the usual (universal) coupling constant are the same if their RG based equations are calculated under the same choice of scheme parameters. This idea has later been discussed in detail by Refs.~\cite{gruta1,gruta2}. Such an effective coupling constant can be used as a reference to define the renormalization procedure, such as $MS$ scheme, $\overline{MS}$ scheme, etc.. 

The RG-invariance states that a physical quantity should be independent of the renormalization scale and renormalization scheme~\cite{callan,symanzik,peter1,peter2,peter3}. This shows that if the effective coupling constant $a(\tau_{\cal R},\{c^{\cal R}_i\})$ corresponds to a physical observable, it should be independent of any other scale $\tau_{\cal S}$ and any other scheme parameters $\{c^{\cal S}_j\}$; i.e.
\begin{eqnarray}
\frac{\partial a(\tau_{\cal R},\{c^{\cal R}_i\})}{\partial \tau_{\cal S}} &\equiv& 0 \;\;\;\;\;{\rm [scale\;\; invariance]} \;,\label{inv-scale} \\
\frac{\partial a(\tau_{\cal R},\{c^{\cal R}_i\})}{\partial c^{\cal S}_j} &\equiv& 0 \, \;\;\;{\rm [scheme\;\; invariance]} \;.\label{inv-sch}
\end{eqnarray}

\noindent{\it Demonstration}: We provide an intuitive demonstration for the RG invariance from the above RG based equations. Given two effective coupling constants $a(\tau_{\cal R},\{c^{\cal R}_i\})$ and $a(\tau_{\cal S},\{c^{\cal S}_i\})$ defined under two different schemes ${\cal R}$ and ${\cal S}$, one can expand $a(\tau_{\cal R},\{c^{\cal R}_i\})$ in a power series of $a(\tau_{\cal S},\{c^{\cal S}_i\})$ through a Taylor expansion:
\begin{widetext}
\begin{eqnarray}
a(\tau_{\cal R},\{c^{\cal R}_i\})&=& a(\tau_{\cal S}+\bar{\tau},\{c^{\cal S}_i +\bar{c}_i\})\nonumber \\
&=& a(\tau_{\cal S},\{c^{\cal S}_i \})+ \left(\frac{\partial a}{\partial \tau}\right)_{\cal S} \bar{\tau} +\sum_{i} \left(\frac{\partial a}{\partial c_i}\right)_{\cal S} \bar{c_i} \nonumber\\
&& +\frac{1}{2!}\left[\left(\frac{\partial^2 a}{\partial \tau^2}\right)_{\cal S} \bar{\tau}^2 +2\left(\frac{\partial^2 a}{\partial\tau \partial c_i}\right)_{\cal S} \bar{\tau}\bar{c}_{i} +\sum_{i,j}\left(\frac{\partial^2 a}{\partial c_{i} \partial c_j}\right)_{\cal S} \bar{c}_{i}\bar{c}_{j} \right] +\frac{1}{3!}\left[\left(\frac{\partial^3 a}{\partial \tau^3}\right)_{\cal S} \bar{\tau}^3 +\cdots \right]+\cdots , \label{asexpand}
\end{eqnarray}
\end{widetext}
where $\bar{\tau}=\tau_{\cal R} -\tau_{\cal S}$, $\bar{c}_i =c^{\cal R}_i - c^{\cal S}_i$ and the subscript ${\cal S}$ next to the partial derivatives means they are evaluated at the point $(\tau_{\cal S},\{c^{\cal S}_i \})$.

The right-hand side of Eq.(\ref{asexpand}) can be regrouped according to the different orders of scheme-parameters $\{\bar{c}_i\}$. After differentiating both side of Eq.(\ref{asexpand}) over $\tau_{\cal S}$, we obtain
\begin{widetext}
\begin{eqnarray}
\frac{\partial a(\tau_{\cal R},\{c^{\cal R}_i\})}{\partial \tau_{\cal S}} =\frac{\partial^{(n+1)} a(\tau_{\cal S},\{c^{\cal S}_i \})} {\partial\tau_{\cal S}^{(n+1)}}\frac{\bar{\tau}^{n}}{n!} + \sum_{i} \frac{\partial^{(n+1)}a(\tau_{\cal S},\{c^{\cal S}_i \})} {{\partial c^{\cal S}_i}\partial\tau_{\cal S}^{(n)}}\frac{\bar{\tau}^{n-1}\bar{c}_{i}}{(n-1)!} + \cdots , \label{rgi-use}
\end{eqnarray}
\end{widetext}
where $n$ stands for the highest perturbative order for a fixed-order calculation. It is noted that Eq.(\ref{rgi-use}) can be further simplified with the help of RG equations (\ref{scale},\ref{scheme}). If we set $n\to\infty$, the right-hand-side of Eq.(\ref{rgi-use}) tends to zero, and we obtain the scale-invariance equation (\ref{inv-scale}). This shows that if $a(\tau_{\cal R},\{c^{\cal R}_i\})$ corresponds to a physical observable (corresponding to the case of $n\to\infty$), it will be independent of any other scale $\tau_S$. Similarly, doing the first derivative of $a(\tau_{\cal R},\{c^{\cal R}_i\})$ with respect to the scheme-parameter $c^{\cal S}_j$, one can obtain the scheme-invariance equation (\ref{inv-sch}).

In another words, if one uses an effective coupling constant $a(\tau_{\cal S},\{c^{\cal S}_i\})$ under the renormalization scheme ${\cal S}$ and with an initial renormalization scale $\{\tau_{\cal S}\}$ to predict the value of $a(\tau_{\cal R},\{c^{\cal R}_i\})$, the RG-invariances (\ref{inv-scale},\ref{inv-sch}) tell us that

\begin{itemize}
\item if we have summed all types of $c^{\cal S}_i$-terms (or equivalently the $\{\beta^{\cal S}_i\}$-terms) into the coupling constant, as is the case of an infinite-order calculation, then our final prediction of $a(\tau_{\cal R},\{c^{\cal R}_i\})$ will be independent of any choice of initial scale $\tau_{\cal S}$ and renormalization-scheme ${\cal S}$.

\item According to Eq.(\ref{rgi-use}), for a fixed-order estimation (i.e. $n\neq \infty $), there is some residual initial-scale dependence. This is reasonable: as shown by Eq.(\ref{asexpand}), for a fixed-order calculation, the unknown $\{\beta^{\cal S}_i\}$-terms in the higher orders are necessary to cancel the scale dependence from the lower-order terms.

 If we can find a proper way to sum up all the known-type of $\{\beta^{\cal S}_i\}$-terms into the coupling constant, and at the same time suppressing the contributions from those unknown-type of $\{\beta^{\cal S}_i\}$-terms effectively, such residual initial scale dependence can be greatly suppressed. The PMC has been designed for such purpose~\cite{pmc1,pmc2,pmc5}, whose properties will be discussed in more detail in the following sections.

\item If setting all the differences of the renormalization scheme parameters, $\bar{c}_i \equiv 0$ ($i=1,2,\cdots$), Eq.(\ref{asexpand}) returns to a scale-expansion series for the coupling constant expanding over itself but specified at another scale; i.e.
\begin{widetext}
\begin{eqnarray}
a(\tau_{\cal R},\{c^{\cal R}_i\})&=& a(\tau_{\cal S},\{c^{\cal R}_i \})+ \left(\frac{\partial a(\tau_{\cal S},\{c^{\cal R}_i \})}{\partial \tau_{\cal S}}\right) \bar{\tau} + \frac{1}{2!}\left(\frac{\partial^2 a(\tau_{\cal S},\{c^{\cal R}_i \})}{\partial \tau_{\cal S}^2}\right) \bar{\tau}^2 +\frac{1}{3!}\left(\frac{\partial^3 a(\tau_{\cal S},\{c^{\cal R}_i \})}{\partial \tau_{\cal S}^3}\right) \bar{\tau}^3 +\cdots . \label{betaseries}
\end{eqnarray}
\end{widetext}
 Using the RG scale-equation (\ref{scale}), the right-hand-side of the above equation can be rewritten as perturbative series of $a(\tau_{\cal S},\{c^{\cal R}_i \})$, whose coefficient at each order is a $\{\beta^{\cal R}_i\}$-series.
\end{itemize}

If one considers $N_c$ to be an analytic variable, then the scale-setting known from the non-Abelian theory $SU(N_c)$ must agree with the Abelian QED theory at $N_c\to 0$. This shows that above discussions are also suitable for QED; i.e. by taking the limit $N_c \to 0$ at fixed $\alpha=C_F \alpha_s$ with $C_F=(N_c^2-1)/2N_c$, we effectively return to the QED case \cite{qed1,qed2}.

\section{self-consistency requirements for a scale-setting method}
\label{sec:self}

It has been noted that if one knows how to set the optimal scale in all cases, then one can translate the result freely from one scheme to another scheme through proper scale relations~\cite{dse1,dse2}. This observation has later been emphasized in Ref.~\cite{{sr}}, where the scale transformation among different schemes  are called ``commensurate scale relations" (CSRs). It shows that even though the expansion coefficients could be different under different renormalization schemes, after a proper scale-setting, one can find a relation between the effective renormalization scales which ensures the total result remain the same under any renormalization schemes. For simplicity, following the suggestion of Ref.\cite{self}, we omit the scheme parameters in the coupling constant in discussing the self-consistent requirements for a scale-setting method, but will retrieve them when necessary.

In principle, the correctness of a scale-setting method can be judged by experimental data. However, it has been suggested that some self-consistency requirements can shed light on the reliability of the scale-setting method~\cite{self}, in which some initial discussions have been presented. These requirements together with their explanations are listed in the following:

\begin{enumerate}
\item { {\bf Existence} and {\bf Uniqueness} of the renormalization scale $\mu$.} Any scale-setting method must satisfy these two requirements.

\item {\bf Reflexivity}. Given an effective coupling $\alpha_s(\mu)$ specified at a renormalization scale $\mu$, we can express it in terms of itself but specified at another renormalization scale $\mu'$,
 \begin{equation}\label{asexp}
 \quad\quad\quad \alpha_s(\mu)=\alpha_s(\mu')+f_1(\mu,\mu')\alpha_s^2(\mu')+ \cdots,
 \end{equation}
 where $f_1(\mu,\mu')\propto\ln(\mu^2/\mu^{'2})$. When the scale $\mu'$ is chosen to be $\mu$, the above equation reduces to a trivial identity.

 From the scale-invariance (\ref{inv-scale}), up to infinite orders, we have
 \begin{equation}\label{start}
 \frac{\partial\alpha_s(\mu)}{\partial\ln{\mu^{'2}}} \equiv 0 .
 \end{equation}
 This, inversely, means that if $\alpha_s(\mu)$ is known (say, a experimentally measured effective coupling), and we try to use the above perturbative equation to ``predict" $\alpha_s(\mu)$ from itself, then any deviation of $\mu'$ from $\mu$ would lead to an inaccurate result due to the truncation of expansion series.

 More explicitly, for a fixed-order expansion with the highest perturbative-order $n$, from Eq.(\ref{rgi-use}), we obtain
 \begin{displaymath}
 \quad\quad \frac{\partial\alpha_s(\mu)}{\partial\ln{\mu^{'2}}} \propto \frac{\left(\ln{\mu^2}/{\mu^{'2}}\right)^{n}}{n!} \frac{\partial^{(n+1)}\alpha_s(\mu') } {\partial(\ln\mu^{'2})^{(n+1)}} .
 \end{displaymath}
 This shows, generally, the right-hand-side of Eq.(\ref{asexp}) depends on $\mu'$ at any fixed-order.

 Thus, to get a correct fixed-order estimate for $\alpha_s(\mu)$, a self-consistency scale-setting must take the unique value ${\mu'} = \mu$ on the right-hand-side of Eq.(\ref{asexp}). If a scale-setting satisfies such property, we say it is {\bf reflexive}.

 It is found that the {\bf Reflexivity} is a basic requirement for a self-consistency scale-setting method and for the physical (effective) coupling constant $\alpha_s(\mu)$, which provides the necessary condition for the following two properties {\bf Symmetry} and {\bf Transitivity}; i.e. if a scale-setting does not satisfy the {\bf Reflexivity}, it cannot satisfy the following two properties {\bf Symmetry} and {\bf Transitivity} either.

\item {\bf Symmetry}. Given two different effective coupling constants $\alpha_{s1}(\mu_1)$ and $\alpha_{s2}(\mu_2)$ under two different renormalization schemes, we can expand any one of them in terms of the other:
 \begin{eqnarray}
\quad\alpha_{s1}(\mu_1)&=&\alpha_{s2}(\mu_2)+ r_{12}(\mu_1,\mu_2)\alpha_{s2}^2(\mu_2)+\cdots,\nonumber\\
\quad\alpha_{s2}(\mu_2)&=&\alpha_{s1}(\mu_1)+ r_{21}(\mu_2,\mu_1)\alpha_{s1}^2(\mu_1)+\cdots.\nonumber
 \end{eqnarray}
 After a general scale-setting, we have
 \begin{eqnarray}
\quad\alpha_{s1}(\mu_1)&=&\alpha_{s2}(\mu^{*}_2)+ \tilde{r}_{12}(\mu_1,\mu^{*}_2) \alpha_{s2}^2(\mu^{*}_2)+\cdots \,, \nonumber\\
\quad\alpha_{s2}(\mu_2)&=&\alpha_{s1}(\mu^{*}_1)+ \tilde{r}_{21}(\mu_2,\mu^{*}_1) \alpha_{s1}^2(\mu^{*}_1)+\cdots \,. \nonumber
 \end{eqnarray}
 Note that,
 \begin{itemize}
 \item The new effective scales $\mu^{*}_{1,2}$ may or may not be equal to $\mu_{1,2}$, depending on the choice of the scale-setting method. The coefficients $\tilde{r}_{12}$ and $\tilde{r}_{21}$ are changed accordingly in order to obtain a consistent result.

 \item We have implicitly set the effective scales at NLO-level to be equal to the LO ones. We will adopt this choice throughout the paper. The effective scales for the highest-order terms are usually taken as the same effective scales at the one-lower-order, since they are the scales strictly set by the known-terms~\cite{HJLu,pmc2}.
 \end{itemize}

 Setting $\mu^{*}_{2}=\lambda_{21}\mu_{1}$ and $\mu^{*}_1= \lambda_{12}\mu_{2}$, if
 \begin{equation}\label{sysbas}
 \lambda_{12}\lambda_{21}=1 \ ,
 \end{equation}
 we say that the scale-setting is {\bf symmetric}. \\

 {\it Explanation}: \\

 If $\mu^{*}_{2}=\lambda_{21}\mu_{1}$ and $\mu^{*}_1= \lambda_{12}\mu_{2}$, we obtain
 \begin{eqnarray}
 &&\alpha_{s1}(\mu_1)\nonumber\\
 &=&\alpha_{s2}(\lambda_{21}\mu_{1})+
 \tilde{r}_{12}(\mu_1,\lambda_{21}\mu_{1})\alpha_{s2}^2(\lambda_{21}\mu_{1})+\cdots
 \label{eq15}
 \end{eqnarray}
 and
 \begin{eqnarray}
 &&\alpha_{s2}(\mu_2)\nonumber\\
 &=&\alpha_{s1}(\lambda_{12}\mu_{2})+
 \tilde{r}_{21}(\mu_2,\lambda_{12}\mu_{2})\alpha_{s1}^2(\lambda_{12}\mu_{2})+\cdots . \label{eq16}
 \end{eqnarray}
 As a combination of Eqs.(\ref{eq15},\ref{eq16}), we obtain
 \begin{widetext}
 \begin{equation} \label{symmetry}
 \alpha_{s1}(\mu_1) = \alpha_{s1}(\lambda_{12}\lambda_{21}\mu_{1})+ \left[
 \tilde{r}_{12}(\mu_1,\lambda_{21}\mu_{1}) + \tilde{r}_{21}(\lambda_{21}\mu_1, \lambda_{12} \lambda_{21}\mu_{1})\right] \alpha_{s1}^2(\lambda_{12} \lambda_{21}\mu_{1})+ \cdots.
 \end{equation}
 \end{widetext}
 From the {\bf Reflexivity} property, if a scale-setting is symmetric, i.e. satisfying Eq.(\ref{sysbas}), we will obtain
 \begin{equation} \label{sysrel}
 \tilde{r}_{12} (\mu_{1},\mu^*_{2}) + \tilde{r}_{21}(\mu_{2},\mu^*_{1})= 0 ,
 \end{equation}
 and vice versa. This shows that the {\bf Symmetry} property (\ref{sysbas}) and the relation (\ref{sysrel}) are mutually necessary and sufficient conditions.

 The {\bf Symmetry} feature is necessary since it further gives us a unique relation for the scales before and after the scale-setting,
 \begin{displaymath}
 \mu_1 \mu_2=\mu_1^* \mu_2^* \,.
 \end{displaymath}

\item {\bf Transitivity}. Given three effective coupling constants $\alpha_{s1}(\mu_1)$, $\alpha_{s2}(\mu_2)$, and $\alpha_{s3}(\mu_3)$ under three renormalization schemes, we can expand any one of them in terms of the other; i.e.
 \begin{eqnarray}
 \quad\alpha_{s1}(\mu_1)&=&\alpha_{s2}(\mu_2)+ r_{12}(\mu_1,\mu_2)\alpha_{s2}^2(\mu_2)
 +\cdots , \nonumber\\
 \quad\alpha_{s2}(\mu_2)&=&\alpha_{s3}(\mu_3)+ r_{23}(\mu_2,\mu_3)\alpha_{s3}^2(\mu_3)
 +\cdots ,\nonumber\\
 \quad\alpha_{s3}(\mu_3)&=&\alpha_{s1}(\mu_1)+ r_{31}(\mu_3,\mu_1)\alpha_{s1}^2(\mu_1)
 + \cdots .\nonumber
 \end{eqnarray}
 After a general scale-setting, we obtain
 \begin{eqnarray}
\quad\alpha_{s1}(\mu_1)&=&\alpha_{s2}(\mu^{*}_2)+ \tilde{r}_{12}(\mu_1,\mu^{*}_2) \alpha_{s2}^2(\mu^{*}_2)+\cdots , \nonumber\\
\quad\alpha_{s2}(\mu_2)&=&\alpha_{s3}(\mu^{*}_3)+ \tilde{r}_{23}(\mu_2,\mu^{*}_3) \alpha_{s3}^2(\mu^{*}_3)+\cdots , \nonumber\\
\quad\alpha_{s3}(\mu_3)&=&\alpha_{s1}(\mu^{*}_1)+ \tilde{r}_{13}(\mu_3,\mu^{*}_1) \alpha_{s1}^2(\mu^{*}_1)+\cdots . \nonumber
 \end{eqnarray}
 Setting $\mu^{*}_2 =\lambda_{21}\mu_1$, $\mu^{*}_3 =\lambda_{32}\mu_2$ and $ \mu^{*}_1 = \lambda_{13}\mu_3$, if
 \begin{equation}\label{transbas}
 \lambda_{13}\lambda_{32}\lambda_{21} =1 \ .
 \end{equation}
 we say that the scale-setting is {\bf transitive}. \\

 {\it Explanation}: \\

 If $\mu^{*}_2 =\lambda_{21}\mu_1$, $\mu^{*}_3 =\lambda_{32}\mu_2$ and $ \mu^{*}_1 = \lambda_{13}\mu_3$, we obtain
 \begin{eqnarray}
 &&\alpha_{s1}(\mu_1)\nonumber\\
 &=&\alpha_{s2}(\lambda_{21}\mu_{1})+
 \tilde{r}_{12}(\mu_1,\lambda_{21}\mu_{1})\alpha_{s2}^2(\lambda_{21}\mu_{1})+\cdots,\\
 \label{eq20}
 &&\alpha_{s2}(\mu_2)\nonumber\\
 &=&\alpha_{s3}(\lambda_{32}\mu_{2})+
 \tilde{r}_{23}(\mu_2,\lambda_{32}\mu_{2})\alpha_{s3}^2(\lambda_{32}\mu_{2})+\cdots , \label{eq21}\\
 &&\alpha_{s3}(\mu_3)\nonumber\\
 &=&\alpha_{s1}(\lambda_{13}\mu_{3})+
 \tilde{r}_{31}(\mu_3,\lambda_{13}\mu_{3})\alpha_{s1}^2(\lambda_{13}\mu_{3})+\cdots . \label{eq22}
 \end{eqnarray}
 As a combination of Eqs.(\ref{eq20},\ref{eq21},\ref{eq22}), we obtain
 \begin{widetext}
 \begin{eqnarray}
 \alpha_{s1}(\mu_1)&=&\alpha_{s1}(\lambda_{13}\lambda_{32}\lambda_{21}\mu_{1})+
 \alpha_{s1}^2(\lambda_{13}\lambda_{32}\lambda_{21}\mu_{1})\times \nonumber\\
 &&\left[ \tilde{r}_{31}(\lambda_{32}\lambda_{21}\mu_{1},
 \lambda_{13}\lambda_{32}\lambda_{21}\mu_1)+
 \tilde{r}_{23}(\lambda_{21}\mu_{1},\lambda_{32}\lambda_{21}\mu_1)+ \tilde{r}_{12}(\mu_{1},
 \lambda_{21}\mu_1)\right]+\cdots . \label{transiv}
 \end{eqnarray}
 \end{widetext}

 From the {\bf Reflexivity} property, if a scale-setting is transitive, i.e. satisfying Eq.(\ref{transbas}), we will obtain
 \begin{equation} \label{transrel}
 \quad\tilde{r}_{12}(\mu_{1},\mu^*_{2}) + \tilde{r}_{23}(\mu^*_{2},\mu^*_{3}) + \tilde{r}_{31}(\mu^*_{3},\mu_{1})= 0 ,
 \end{equation}
 and vice versa. This shows that the {\bf Transitivity} property (\ref{transbas}) and the relation (\ref{transrel}) are mutually necessary and sufficient conditions.

 The {\bf Transitivity} property shows that under a proper scale-setting method, we have $\lambda_{21}\equiv\lambda_{23}\lambda_{31}$, which means that the scale ratio $\lambda_{21}$ for any two effective couplings $\alpha_{s1}$ and $\alpha_{s2}$ is independent of the choice of an intermediate effective coupling $\alpha_{s3}$ under any renormalization scheme. Thus the relation between any two observables is independent of the choice of renormalization scheme. In fact, the {\bf Transitivity} property provides the theoretical foundation for the existence of CSRs among different physical observables~\cite{sr}.

 The {\bf Transitivity} feature gives us a unique relation for all the scales before and after the scale-setting,
 \begin{displaymath}
 \mu_1 \mu_2 \mu_3 = \mu_1^* \mu_2^* \mu_3^* \,.
 \end{displaymath}

 The {\bf Transitivity} property is very important for a self-consistency scale-setting, which is a natural requirement from the RG invariance. It has already been pointed out that why the renormalization group is called a ``group" is mainly because of such {\bf Transitivity} property \cite{peter1,peter2,peter3}.

 The {\bf Transitivity} property (\ref{transbas}) can be extended to an arbitrary number of effective coupling constants; i.e. if we have $n_{th}$ effective coupling constants which are related with similar manner as above, then their transitivity relation is
 \begin{equation}
 \lambda_{1n}\lambda_{n(n-1)}\cdots\lambda_{32}\lambda_{21} =1 .
 \end{equation}

 One may observe that the {\bf Symmetry} is a special case of {\bf Transitivity}, since if setting $\alpha_{s3}(\mu_3)\equiv\alpha_{s1}(\mu_1)$, we have $\lambda_{11} \equiv 1$ and $\tilde{r}_{11}(\mu_{1},\mu_{1}) \equiv 0$ due to the {\bf Reflexivity}, which thus changes the transitive relation $\lambda_{13}\lambda_{32} \lambda_{21}=1$ into the symmetric relation $\lambda_{12}\lambda_{21}=1$.

\end{enumerate}

As a summary, a scale-setting method that satisfies {\bf Existence} and {\bf Uniqueness} of the renormalization scale, {\bf Reflexivity}, {\bf Symmetry}, and {\bf Transitivity} effectively establishes equivalent relations among all the effective coupling constants, and thus, among all the physical observables.

\subsection{A graphic explanation of these requirements}

In this subsection, we present a more intuitive explanation of these requirements based on the universal coupling $a(\tau,\{ c_i \})$ and the RG based Eqs.(\ref{scale},\ref{scheme}).

In the RG based equations (\ref{scale},\ref{scheme}), there is no explicit reference to the QCD parameters, such as the number of colors or the number of active-flavors. Therefore, aside from its infinite dimensional character, $a(\tau,\{ c_i \} )$ is just a mathematical function like, say, Bessel functions or any other special functions~\cite{HJLu}. In practice, due to the unknown higher order scheme parameters $\{ c_i \}$, we need to truncate the beta function $\beta(a,\{ c_i \} )$ and solve the universal coupling constant $a(\tau,\{ c_i \} )$ in a finite-dimensional subspace; i.e. we need to evaluate $a(\tau,\{ c_i \})$ in a subspace where higher order $\{c_i\}$-terms are zero. In principle, this function can be computed to arbitrary degree of precision, limited only by the truncation of the fundamental beta function.

\begin{figure}[ht]
\includegraphics[width=0.40\textwidth]{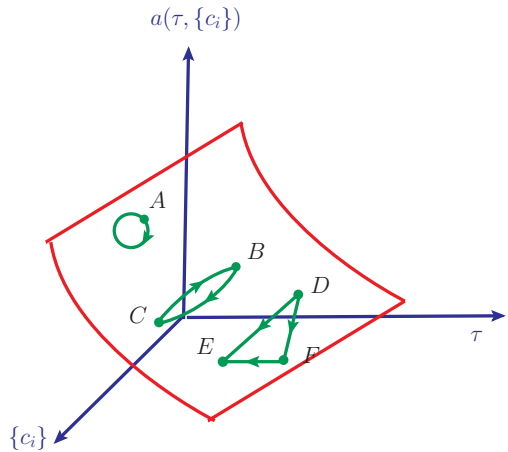}
\caption{Pictorial representation of the self-consistency of the scale-setting method through the universal coupling function $a(\tau,\{c_i\})$. The point $A$ with a closed path represents the operation of {\bf Reflexivity}. The paths $\overline{BC}$ and $\overline{CB}$ represent the operation of {\bf Symmetry}, and the paths $\overline{DF}$,$\overline{FE}$ and $\overline{DE}$ represent the operation of {\bf Transitivity}. }
\label{fig3}
\end{figure}

In this formalism, any two effective coupling constants can be related by some evolution path on the hypersurface defined by $a(\tau,\{ c_i \} )$. In Fig.(\ref{fig3}) we illustrate the paths which represent the operations of {\bf Reflexivity}, {\bf Symmetry} and {\bf Transitivity}. We can pictorially visualize that the evolution paths satisfy all these three self-consistency properties. A closed path starting and ending at the same point $A$ represents the operation of identity. Since the predicted value does not depend on the chosen path, if the effective coupling constant at $A$ is $a_A$, after completing the path we will also end up with an effective coupling $a_A$. Similarly, if we evolve $a_B$ at $B$ to a value $a_C$ at $C$, we are guaranteed that when we evolve $a_C$ at $C$ back to the point $B$, the result will be $a_B$. Hence, the evolution equations also satisfy {\bf Symmetry}. {\bf Transitivity} follows in a similar manner; i.e. going directly from $D$ to $E$ gives the same result as going from $D$ to $E$ through a third point $F$.

In the following two sections, we will make a detailed discussion on how these self-consistency conditions are satisfied or broken by the two frequently adopted scale-setting methods: BLM/PMC and PMS. As for FAC, its FAC scale is determined by requiring all higher order corrections to be zero \footnote{This method itself is useful to define an effective coupling constant for a physical process~\cite{FAC1,FAC2,FAC3}. However it will give wrong result when applied to QED processes. The FAC forces all higher order corrections to vanish and runs the risk of the better approximation being ``dragged down" by the poorer one~\cite{PMS2}.}. FAC satisfies all the above mentioned self-consistenct requirements, whose demonstration is similar to that of BLM/PMC and is simpler~\cite{self}, so we will not repeat it here.

\section{The PMC scale-setting}
\label{sec:pmc}

The PMC provides the principle underlying BLM scale-setting, so if not specially stated, we usually treat them on equal footing.

\subsection{What is PMC ?}

In the original BLM paper~\cite{BLM}, the physical observable is expanded as
\begin{eqnarray}
\rho = C_{0}\alpha_{s,\overline{MS}}(\mu) \left[1+ \left(A n_{f}+B\right) \frac{\alpha_{s,\overline{MS}}(\mu)}{\pi} \right],
\end{eqnarray}
where $\mu$ is the renormalization scale, the $n_f$ term is due to the quark vacuum polarization. For clarity, we have taken the familiar $\overline{MS}$-scheme. When absorbing all the NLO terms involving $n_f$ into the running coupling, we obtain~\cite{BLM}
\begin{equation}
\rho=C_{0}\alpha_{s,\overline{MS}}(\mu^*) \left[1+ C^*_{1} \frac{\alpha_{s,\overline{MS}}(\mu^*)}{\pi} \right], \label{blmorg}
\end{equation}
where
\begin{equation}
\mu^*=\mu \exp\left({3A}\right) \;\;{\rm and}\;\; C_1^*=\frac{33}{2}A +B \;.
\end{equation}
The new scale $\mu^*$ and the coefficient $C_1^*$ are $n_f$ independent. The term $33A/2$ in $C_1^*$ serves to remove that part of the constant $B$ which renormalizes the NLO coupling constant.

Through these procedures, it was suggested that the pQCD convergence can be greatly improved~\cite{BLM}. However, after a proper extension of BLM, it can do much more than that.

In deriving Eq.(\ref{blmorg}), Brodsky-Lepage-Mackenzie already observed that to derive the correct scheme-independent LO QED/QCD scale, one should deal with the $\beta_0$-term rather than the $n_f$-term. This point has lately been emphasized in Refs.~\cite{pomeron,pomeron1}, where an interesting feature for the NLO BFKL Pomeron intercept function $\omega(Q^2,0)$ has been found; i.e. after using BLM scale-setting, the function $\omega(Q^2,0)$ has a very weak dependence on the gluon virtuality $Q^2$ in comparison with that derived from the conventional scale-setting under the MOM scheme and $\overline{MS}$ scheme \cite{pomeron}. The BLM has also been applied with some modifications for determining the effective scale in lattice perturbative theory by Lepage and Mackenzie~\cite{lepage2}, which greatly enhances the predictive power of lattice perturbative theory. However, BLM in its original form is difficult to be applied to higher order calculations because of the emergence of higher order $n_f$-terms as $n_f^2$-term, $n_f^3$-term, etc..

As an extension of BLM scale-setting, a program to deal with higher order $n_f$-terms associated with renormalization has been raised in Ref.~\cite{kataev1}, which suggests that one can expand the effective scale as a perturbative series. Later on, an enhanced discussion of this suggestion up to NNLO level has been presented in Ref.~\cite{sr}, where the perturbative series of the effective scale is exponentiated, which is consistent with PMC procedure. In that work it is pointed out that the $n_f^2$-term at the NNLO should be first identified with $\beta^2_0$-term and then be absorbed into the coupling constant \footnote{Strictly, together with the $n_f$-term at the same order, it should be arranged into a proper linear combination of $\beta_1$-term and $\beta_0^2$-term, and the $\beta_0^2$-term will be absorbed into LO PMC scale and $\beta_1$-term will be absorbed into NLO PMC scale \cite{pmc2}. }.

The pioneering work for PMC is done in Ref.~\cite{pmc1}, which shows that a single global PMC scale, valid at LO, can be derived from basic properties of the perturbative QCD cross-section. Later on, explicit formulae for setting PMC scales up to NNLO has been presented in Ref.~\cite{pmc2}. It has also been pointed out that by introducing the PMC-BLM correspondence principle, we can improve the previous BLM procedure to deal with the process up to all orders, whose estimation is the same as PMC. In this sense, we say that PMC and BLM are equivalent to each other. Recently, by applying PMC to the top-quark pair hadroproduction up to NNLO level at the Tevatron and LHC colliders, the most striking feature of PMC has been observed, which shows that the PMC scales and the resulting finite-order PMC predictions are both to high accuracy independent of the choice of an initial renormalization scale, consistent with RG-invariance~\cite{pmc3,pmc4,pmc5}. This implies that the serious systematic renormalization scale error introduced by using conventional scale-setting can be eliminated by PMC through a self-consistency way.

\begin{figure}
\includegraphics[width=0.40\textwidth]{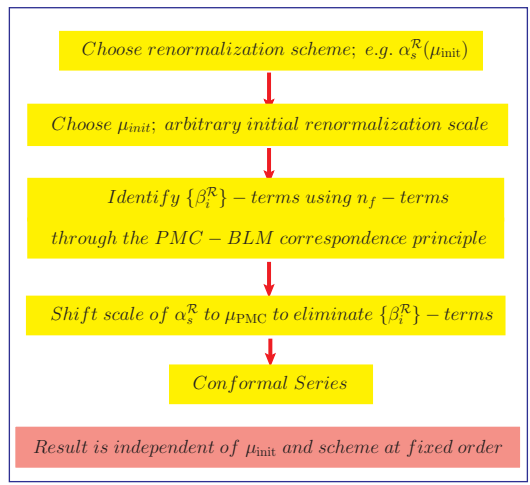}
\caption{A ``flow chart" which illustrates the PMC procedure, where ${\cal R}$ stands for an arbitrary renormalization scheme. }
\label{fig1}
\end{figure}

A ``flow chart" which illustrates the PMC procedure is presented in Fig.(\ref{fig1}), where ${\cal R}$ stands for an arbitrary renormalization scheme. The PMC provides a unambiguous and systematic way to set the optimized renormalization scale up to all orders; i.e. we first arrange all the coefficients, which usually are given as a series in $n_f$, for each perturbative order into $\{\beta^{\cal R}_i\}$-terms or non-$\{\beta^{\cal R}_i\}$-terms, and absorb different types of $\{\beta^{\cal R}_i\}$-term into the running coupling constant, order-by-order \footnote{In practice, we can directly deal with $n_f$-terms of the coefficients without changing them into $\{\beta^{\cal R}_i\}$-terms, and eliminate the $n_f$-terms from the highest power to none also in an order-by-order manner. The results are the same due to the PMC-BLM correspondence~\cite{pmc2}. }. Different types of $\{\beta^{\cal R}_i\}$-term are absorbed into different PMC scales. Different skeleton graphs can have different PMC scales. The PMC scales themselves will be a perturbative expansion series in $\alpha_s$. After this procedure, all non-conformal $\{\beta^{\cal R}_i\}$-terms in the perturbative expansion are resummed into the running couplings so that the remaining terms in the perturbative series are identical to that of a conformal theory; i.e., the corresponding theory with $\{\beta^{\cal R}_i\} \equiv \{0\}$.

As a simple explanation of PMC, for the coefficient ${\cal C}_1(\mu)$ at the NLO level, we have
\begin{eqnarray}
{\cal C}_1(\mu) &=& {\cal C}_{10}(\mu) + {\cal C}_{11}(\mu) n_f, \\
 &=& \tilde{{\cal C}}_{10}(\mu) + \tilde{{\cal C}}_{11}(\mu) \beta_0
\end{eqnarray}
where $\mu$ stands for an arbitrary initial renormalization scale, the coefficients ${\cal C}_{10}(\mu)$ and ${\cal C}_{11}(\mu)$ are $n_f$-independent, $\tilde{{\cal C}}_{10}={\cal C}_{10}+\frac{33}{2}{\cal C}_{11}$, and $\tilde{{\cal C}}_{11}=-\frac{3}{2}{\cal C}_{11}$. The LO PMC scale $\mu_{\rm PMC}$ is then set by the condition
\begin{equation}\label{pmcbasic}
\tilde{{\cal C}}_{11}(\mu_{\rm PMC}) = 0.
\end{equation}
This prescription ensures that, as in QED, vacuum polarization contributions due to the light-fermion pairs are absorbed into the coupling constant. Note that because ${{\cal C}}_{11}\propto\tilde{{\cal C}}_{11}$, one can practically obtain the PMC scale by using the equation ${{\cal C}}_{11}(\mu_{\rm PMC}) \equiv 0$, which is usually adopted in the literature \footnote{This should be used with care, since if ${\cal C}_{10}$ is a constant free of scale, then such practical way will give wrong NLO coefficient other than the correct one ${\tilde{\cal C}}_{10}$. }. However one should keep in mind that Eq.(\ref{pmcbasic}) is exact.

The PMC - BLM correspondence principle suggested in Ref.~\cite{pmc2} is based on the fact that the purpose of the running coupling in any gauge theory is to sum up all the terms involving the $\{\beta^{\cal R}_i\}$-functions; conversely, one can find all the needed $\{\beta^{\cal R}_i\}$-terms at any relevant order by identifying terms arising from the order-by-order expansion of the running coupling. This principle provides a one-to-one correspondence between the $n_f$-series and the $\beta^{\cal R}_i$-series, and it provide a practical way of identifying the terms in the $n_f$-series into the required $\beta^{\cal R}_i$-series. The $\{\beta^{\cal R}_i\}$-series derived from Eq.(\ref{betaseries}) provides the foundation for the PMC - BLM correspondence principle, since it shows which $\{\beta^{\cal R}_i\}$-terms should be kept at a specific perturbative order.
This procedure provides a convenient and consistent way of treating the $\{\beta^{\cal R}_i\}$-terms in the perturbative series. Its advantages will be shown in the next subsection. Such a choice of $\{\beta^{\cal R}_i\}$-series is not completely identical to the suggestion of Refs.~\cite{kataev1,kataev2,kataev3,kataev4}. In Refs.~\cite{kataev1,kataev2,kataev3}, as an extension of BLM scale-setting to all orders, the large $\beta_0$-approximation is adopted with some modifications to simplify the calculation (called as the seBLM and the $x$BLM approach~\cite{kataev2}) \footnote{Theoretical differences for different treatments will be discussed in more detail and will be presented elsewhere.}.

\subsection{The properties of PMC}

It is straightforward to verify that PMC satisfies all the self-consistency requirements outlined above.

\begin{enumerate}

\item The {\bf Existence} and {\bf Uniqueness} of the renormalization scale $\mu$ are guaranteed, since the scale-setting conditions for PMC are often linear equations in $\ln \mu^2$.

 As a simpler explanation, if the NLO coefficient ${\cal C}_1(\mu)$ in Eq.(\ref{phyvalue}) has the form
 \begin{equation}
 {\cal C}_1(\mu) = ( a + b\; n_f) + ( c + d\; n_f) \ln\mu^2 ,
 \end{equation}
 with $a, b, c$ and $d$ are constants free of $n_f$. The LO PMC scale can be set as
 \begin{equation}
 \ln{\mu_{\rm PMC}^{LO}} = - \frac{b}{2d} + {\cal O}(\alpha_s) ,
 \end{equation}
 where the higher order $\alpha_s$-terms will be determined by $n_f$-terms at the NLO-level or even higher levels.

\item {\bf Reflexivity} is satisfied. The PMC requires all $\ln(\mu^2/\mu'^2)$-terms in Eq.(\ref{asexp}) vanish, thus we obtain
 \begin{displaymath}
 \mu'=\mu \,.
 \end{displaymath}

\item {\bf Symmetry} is trivial, because after PMC scale-setting, we always have
 \begin{displaymath}
 \tilde{r}_{12}(\mu_{1},\mu^*_{2}) =- \tilde{r}_{21}(\mu_{2},\mu^*_{1})\,.
 \end{displaymath}
 That is, the two NLO coefficients only differ by a sign. Thus, requiring one of them to be $\{\beta^{\cal R}_i\}$-independent is equivalent to requiring the other one also to be $\{\beta^{\cal R}_i\}$-independent. This argument ensures the symmetric relation, $\lambda_{12}\lambda_{21}=1$, is satisfied after PMC scale-setting.

\item {\bf Transitivity} is also satisfied by PMC. After PMC scale-setting, the two coefficients $\tilde{r}_{12}(\mu_1,\mu^*_2)$ and $\tilde{r}_{23}(\mu^{*}_2,\mu^*_3)$ in the following two series
 \begin{eqnarray}
 &&\alpha_{s1}(\mu_1) \nonumber\\
 &=& \alpha_{s2}(\mu^*_2) + \tilde{r}_{12}(\mu_1,\mu^*_2) \alpha_{s2}^2(\mu^*_2)+ {\cal O}(\alpha_{s2}^3) \label{Eq12}
 \end{eqnarray}
 and
 \begin{eqnarray}
 &&\alpha_{s2}(\mu^*_2) \nonumber\\
 &=& \alpha_{s3}(\mu^*_3) + \tilde{r}_{23}(\mu^*_2,\mu^*_3) \alpha_{s3}^2(\mu^*_3)+ {\cal O}(\alpha_{s3}^3) \,, \label{Eq23}
 \end{eqnarray}
 should be independent of $\{\beta_i\}$. After substituting Eq.(\ref{Eq23}) into Eq.(\ref{Eq12}), we obtain
 \begin{eqnarray}
 &&\alpha_{s1}(\mu_1) \nonumber\\
 &=& \alpha_{s3}(\mu^*_3) + [\tilde{r}_{12}(\mu_1,\mu^*_2) + \tilde{r}_{23}(\mu^*_2,\mu^*_3)]\alpha_{s3}^2(\mu_3) \nonumber\\
 && \quad\quad\quad + {\cal O}(\alpha_{s3}^3) \,.
 \end{eqnarray}
 We see that the new NLO coefficient $[\tilde{r}_{12}(\mu_1,\mu^*_2) + \tilde{r}_{23}(\mu^*_2,\mu^*_3)]$ will also be $\{\beta^{\cal R}_i\}$-independent, since it is the sum of two $\{\beta^{\cal R}_i\}$-independent quantities. These arguments ensure the transitive relation, $\lambda_{31}=\lambda_{32}\lambda_{21}$, be satisfied after PMC scale-setting.

\end{enumerate}

As a combination of all the above mentioned PMC features, the advantages of PMC are clear \footnote{In the PMC, the same procedure is valid for both space-like and time-like arguments; in particular this leads to well-behaved perturbative expansion, since all the large $\{\beta^{\cal R}_i\}$-dependent terms on the time-like side involving $\pi^{2}$-terms are fully absorbed into the coupling. The PMC does not change the space-like or time-like nature of the initial renormalization scale $Q_0$, since in general, all the PMC scales are equal to $Q_0$ times an exponential factor~\cite{pmc2}. }:

\begin{itemize}

\item It keeps the information of the higher order corrections but in a more convergent perturbative series. After PMC scale-setting, the divergent ``renormalon" series with $n!$-growth disappear in the perturbative series, so that a more convergent perturbative series is obtained. Such better convergence has already been found in the original BLM paper~\cite{BLM} and the following BLM-literature even at the NLO level.

\item After PMC scale-setting, the renormalization scale dependence is transformed to the initial renormalization scale dependence, and it is found that such initial renormalization scale dependence can be highly suppressed or even eliminated :
 \begin{itemize}
 \item The resulting expressions are conformally invariant and thus do not depend on the choice of renormalization scheme.

 \item One can obtain proper scale-displacements among the PMC scales which are derived under different schemes or conventions.

 \item One can obtain the CSR between any two physical observables such as the Generalized Crewther Relation connecting the Bjorken sum rule to the $e^+ e^-$ annihilation cross section. Many leading order CSRs have been derived in Ref.\cite{sr}. The CSRs have no scale ambiguity and are independent of the choice of renormalization scheme. The relative scales in the CSR ensure that two observables pass through each quark threshold in synchrony. The coefficients in the CSR can be identified with those obtained in conformally invariant gauge theory~\cite{kataev3,crewther1,crewther2,crewther3,crewther4}.

 \item There can be some residual scheme dependence for a fixed-order calculation due to unknown higher-order terms. However this scheme-dependence can be highly suppressed in a similar way as that of the residual initial scale dependence; such effects can be estimated by using the RG-based scheme equations~\cite{HJLu,pmc2}.

 \end{itemize}

\item The PMC provides a fundamental and systematic way to set the optimized renormalization scale for the fixed-order calculation. In principle, PMC needs an initial renormalization scale to initialize it. However, it is found that the estimates after PMC scale-setting are independent of any choice of the initial renormalization scale - even the PMC scales themselves are independent of any choice of initial scale and are `physical' at any fixed order. This is because that the PMC scale itself is a perturbative series and the unknown higher-order $\{\beta^{\cal R}_i\}$-terms are to be absorbed into the higher-order term of PMC scale and will be strongly power suppressed. One example of this behavior is shown in Refs.~\cite{pmc3,pmc4,pmc5}, where the top-quark pair total cross-section and the top-quark pair forward-backward asymmetry are almost free from the choice of initial renormalization scale even at the NNLO-level.

\item Moreover, it is found that the PMC scale-setting can also be adopted for QED case. The variable $N_C$ can be taken as an analytic variable. In the Abelian limit $N_C \to 0$ at fixed $\alpha=C_F \alpha_s$ with $C_F=(N_c^2-1)/2N_c$~\cite{qed1}, the PMC method agrees with the standard Gell Mann-Low procedure for setting the renormalization scale in QED, a consistency requirement of analyticity of Yang Mills gauge theories.

\item After PMC scale-setting, the number of active flavors $n_f$ is correctly determined~\cite{msext}. Using the PMC ensures that the expansion is unchanged as one passes each quark threshold, since all vacuum-polarization effects due to each new quark are automatically absorbed into the effective coupling constant.

\item The argument of the running coupling has time-like or space-like values appropriate to the physics of the PMC scale; for example the scale of the QED coupling which some all vacuum polarization corrections in the lowest order $e^+ e^- \to \mu^+ \mu^-$ amplitude is $\alpha(t)$ in the Gell Mann-Low scheme. As in QED, the running QCD coupling is complex in the time-like domain, reflecting the contribution of diagrams with physical unitarity cuts.

\end{itemize}

\section{The PMS scale-setting}
\label{pms}

\subsection{What is PMS ?}

The PMS states that~\cite{PMS1,PMS2,PMS3,PMS4} if an estimate depends on some ``unphysical" parameters \footnote{Here the ``unphysical" parameter means which is known not to affect the true result. }, then their values should be chosen so as to minimize the sensitivity of the estimate to small variations in these parameters; i.e. this method chooses $\mu_{\rm PMS}$ at the stationary point of $\rho_N$ :
\begin{equation}
 \frac{\partial \rho_N}{\partial \mu}\left|_{\mu=\mu_{\rm PMS}} \equiv 0 \right.
\end{equation}
or
\begin{equation}\label{pmsbasic}
 \frac{\partial \rho_N}{\partial \ln\left(\mu^2\right)}\left|_{\mu=\mu_{\rm PMS}} \equiv 0 \right. .
\end{equation}
Here Eq.(\ref{pmsbasic}) can be solved with the help of the usual renormalization group equation (\ref{basic-RG}).

\subsection{The properties of PMS}

Unlike the case of PMC, in general, there are no known theorems that guarantee the {\bf Existence} or the {\bf Uniqueness} of the PMS solution. Although for practical cases PMS does provide solutions, and when there are more than one solution usually only one of them lies in the physically reasonable region~\cite{PMS1,PMS2,PMS3,PMS4}, these observations alone do not guarantee that PMS will be trouble-free for new processes.

To discuss PMS properties in a renormalization scheme-independent way, following the suggestion of Ref.~\cite{self}, we adopt the 't Hooft scheme \cite{tH} to define the running behavior of the effective coupling constant. Under the 't Hooft scheme, all the scheme parameters $\{c_i\}$ are set to zero, and Eq.(\ref{scale}) simplifies to
\begin{equation}
\frac{d a}{d\tau}=-a^2(1+a) ,
\end{equation}
whose solution can be written as
\begin{equation}
\tau =\frac{1}{a} + \ln \left( \frac{a}{1+a} \right).
\end{equation}
In the above solution, for convenience, we have redefined $\tau$ as $\frac{\beta_0^2}{\beta_1} \ln\left(\frac{\mu^2}{\Lambda^{'tH 2}_{QCD}}\right)$, where $\Lambda^{'tH}_{QCD}$ is the asymptotic scale under the 't Hooft scheme. The 't Hooft coupling constant has a formal singularity, $a(\tau,\{c_i\})\equiv a(0,\{0\})=\infty$, which provides a precise definition for the asymptotic scale $\Lambda^{'tH}_{QCD}$~\cite{tH}; i.e. it is defined to be the pole of the coupling function.

Given two effective coupling constants $a_1$ and $a_2$ under the 't Hooft scheme, they are related by the perturbative series
\begin{equation}
a_1(\tau_1) = a_2(\tau_2) + (\tau_2-\tau_1) a_2^2(\tau_2) + \cdots .
\end{equation}
PMS proposes the choice of $\mu_2$ (or equivalently, $\tau_2$) at the stationary point, {\it i.e.}:
\begin{equation}
\frac{d a_1}{d \tau_2} = 0 = \frac{d}{d \tau_2} \left[ a_2(\tau_2) + (\tau_2-\tau_1)
 a_2^2(\tau_2) \right] .
\end{equation}
Then, we obtain the condition:
\begin{equation}
1+a_2 = \frac{1}{2(\tau_1-\tau_2)} .
\end{equation}
In order to obtain $\tau_2$ in terms of $\tau_1$, one must solve the last equation in conjunction with
\begin{equation}
\frac{1}{a_2} + \log \left( \frac{a_2}{1+a_2} \right) =\tau_2 .
\end{equation}

\begin{figure}[ht]
\includegraphics[width=0.40\textwidth]{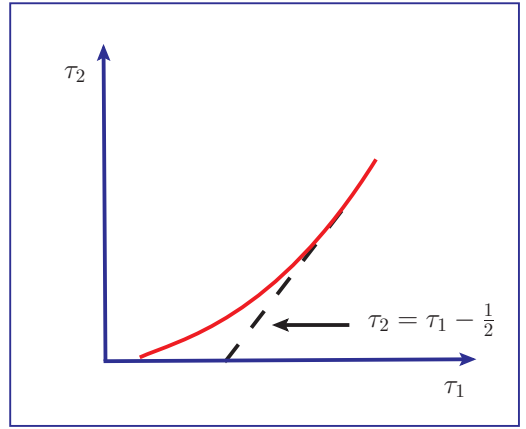}
\caption{The dependence of the PMS scale parameter $\tau_2$ as a function the external scale parameter $\tau_1$. }
\label{fig2}
\end{figure}

In Fig.\ref{fig2} we present the graphical solution of the PMS scale-parameter $\tau_2$ as a function of the external scale-parameter $\tau_1$. One may observe two points:
\begin{itemize}
\item $\tau_2 \geq \tau_1 -\frac{1}{2}$. Since $\tau_2 \neq \tau_1$ in any cases, so PMS explicitly violates the {\bf Reflexivity}. For a fixed-order estimation, when one uses an effective coupling constant to predict itself, the application of PMS would lead to an inaccurate result.

\item In the large momentum region $(\tau_1 \gg 1)$, we obtain $a_2(\tau_2)\to 0$, and
 \begin{equation}\label{pmsrel}
 \tau_2 \simeq \tau_1 - \frac{1}{2} .
 \end{equation}
 Under the same renormalization scheme $R$, we have the same asymptotic parameter $\Lambda^{'tH-R}_{QCD}$ for both $a_1$ and $a_2$. Here $\Lambda^{'tH-R}_{QCD}$ is the 't Hooft scale associated with the $R$-scheme, where the word ``associated" means we are choosing the particular 't Hooft scheme that shares the same 't Hooft scale with the $R$-scheme. Then the relation (\ref{pmsrel}) in terms of $\mu_1$ and $\mu_2$ becomes
 \begin{equation}
 {\mu_2}\simeq {\mu_1} \exp \left( -\frac{\beta_1}{4 \beta_0^2}\right ) .
 \end{equation}
\end{itemize}

More generally, it is found that after PMS scale-setting, the scale displacement between any two scales $\mu_i$ and $\mu_j$ in the large momentum region is
\begin{equation}
\lambda_{ij} = \frac{\mu_i}{\mu_j} \simeq \exp\left(-\frac{\beta_1}{4\beta_0^2}\right) .
\end{equation}
This would mean that
\begin{eqnarray}
\lambda_{12} \lambda_{21} &\simeq& \exp\left(-\frac{\beta_1}{2 \beta_0^2}\right) \neq 1 , \\
\lambda_{13} \lambda_{32} \lambda_{21} &\simeq& \exp\left(-\frac{3\beta_1}{4 \beta_0^2}\right) \neq 1 . \label{pmsneq}
\end{eqnarray}
This shows that the PMS does not satisfy the {\bf Symmetry} and {\bf Transitivity} requirements. Let us point out that adding the scheme-parameter optimization in PMS does not change any of the above conclusions. It only makes the solution much more complicated~\cite{pmsrs}. The inability of PMS to meet these self-consistency requirements resides in that the derivative operations in general do not commute with the operations of {\bf Reflexivity}, {\bf Symmetry} and {\bf Transitivity}.

As argued in Sec.III, any truncated perturbative series will explicitly break RG-invariance (\ref{inv-scale}); i.e. Eq.(\ref{inv-scale}) can only be approximately satisfied for any fixed-order estimation. The precision depends on to which perturbative order we have calculated, the convergence of the perturbative series, and how we set the renormalization scale. As shown by Eq.(\ref{pmsbasic}), the PMS requires the truncated series, i.e. the approximant of a physical observable, to satisfy the RG-invariance near $\mu=\mu_{\rm PMS}$. This provides the underlying reason for why PMS does not satisfy the {\bf Reflexivity}, {\bf Symmetry} and {\bf Transitivity} properties. This shows the necessity of further careful studies of theoretical principles lying below PMS.

The PMC and PMS scale-setting methods each gives specific predictions for physical observables at finite order; however, their predictions are very different:

\begin{itemize}

\item The PMC sums all $\{\beta^{\cal R}_i\}$-terms in an arbitrary renormalization scheme ${\cal R}$ in the fixed-order prediction into the running coupling, leaving the conformal series. It satisfies all of the RG-properties {\bf Reflexivity}, {\bf Symmetry}, and {\bf Transitivity}. The PMC prediction is thus scheme-independent, and it automatically assigns the correct displacement of the intrinsic scales between schemes. The variation of the prediction away from the PMC scale exposes the non-zero $\{\beta^{\cal R}_i\}$-dependent terms. The PMC prediction does have small residual dependence on the initial choice of scale due to the truncated unknown higher order $\{\beta^{\cal R}_i\}$-terms, which will be highly suppressed by proper choice of PMC scales.

\item The PMS chooses the renormalization scale such that the first derivative of the fixed-order calculation with respect to the scale vanishes. This criterion of minimal sensitivity gives predictions which are not the same as the conformal prediction, and the PMS prediction depends on the choice of renormalization scheme \footnote{As shown in Ref.~\cite{pmsrs}, by using the PMS together with the scheme-equations (\ref{scheme}) and the scheme-independent equation (\ref{inv-sch}), such renormalization scheme dependence can be reduced to a certain degree through an order-by-order procedure.}, and it disagrees with QED scale-setting in the Abelian limit. For example, in the case of $e^+ e^- \to g q \bar q$, the PMS scale decreases with increasing gluon jet mass and increasing flavor number, opposite to the correct physical behavior~\cite{Kramer}. The PMS does not satisfy the RG-properties of {\bf Symmetry}, {\bf Reflexivity}, and {\bf Transitivity}, so that relations between observables depend on the choice of the intermediate renormalization scheme.

\end{itemize}

\section{Summary}
\label{sec:summary}

The conventional scale-setting procedure assigns an arbitrary range and an arbitrary systematic error to fixed-order pQCD predictions. As we have discussed in this article, this {\it ad hoc} assignment of the range and associated systematic error is unnecessary and can be eliminated by a proper scale-setting method.

Renormalization group invariance (\ref{inv-scale}) states that a physical quantity should be independent of the renormalization scale and renormalization scheme. With the help of the RG based equations which incorporate the scheme parameters, we have presented a general demonstration for the RG-invariance by setting the perturbative series up to infinite orders.

We have discussed the necessary self-consistency conditions for a scale-setting method, such as the {\bf Existence} and {\bf Uniqueness} of the renormalization scale, {\bf Reflexivity}, {\bf Symmetry}, and {\bf Transitivity}. There properties are natural deductions of RG-invariance. We have shown that PMC satisfies these requirements, whereas the PMS does not. We have also pictorially argued that the formalism based on the RG based equations satisfies all these requirements for scale and scheme variation.

The principle of minimum sensitivity (PMS) requires that the slope of the approximant of an observable to vanish at the renormalization point. With the help of the RG based equations, it has been argued that PMS can provide renormalization-scheme dependent estimates~\cite{PMS1,PMS2,PMS3,PMS4}. We have shown that the PMS violates the {\bf Symmetry} and {\bf Transitivity} properties of the renormalization group, and it does not reproduce the Gell-Mann-Low scale for QED observables. Eq.(\ref{pmsneq}) shows that the relation between any two physical observables after PMS scale-setting depends on which renormalization scheme chosen for the calculation, which explicitly breaks the ``group properties" of the RG equations. In addition, the application of PMS to jet production gives unphysical results~\cite{Kramer}, since it sums physics into the running coupling not associated with renormalization. This implies the necessity of further careful studies of theoretical principles lying below PMS.

The principle of maximum conformality (PMC) provides a fundamental and systematic way to set the optimized renormalization scale at fixed order in pQCD. The PMC has a solid theoretical background~\cite{pmc1,pmc2}, it provides the underlying principle for BLM, and many PMC features have already been noted in the BLM literature. Most important, it is found after standard PMC scale-setting, the theoretical prediction is essentially independent of the choice of initial renormalization scale and the theorist's choice of renormalization scheme, consistent with the RG invariance.

The most important goal for a scale-setting method is to eliminate the renormalization scheme and initial scale dependences -- more fundamental requirements than improving convergence of the pQCD series. In the literature, however, some extensions of BLM scale-setting have concentrated on how to improve the pQCD convergence, such as the large $\beta_0$-expansion~\cite{blm3}, the sequential BLM (seBLM) and xBLM~\cite{kataev1,kataev2,kataev3}, etc.. In fact, once one sets the scales properly, as PMC does, much better pQCD convergence than the conventional scale-setting method is automatic, since the divergent ``renormalon" series with $n!$-growth has been absorbed into the effective scales and disappears in the perturbative series. An example of this improved convergence can be found in our analysis for the top-quark pair production at the NNLO level~\cite{pmc3,pmc4,pmc5}.

Two more subtle points for PMC scale-setting:

\begin{itemize}

\item In some specific kinematical regions, such as for the heavy quark pair production in the threshold region, Coulomb-type corrections will lead to sizable contributions which are enhanced by factors of $\pi/v$ and the PMC scale can be relatively soft for $v\to 0$ ($v$, the heavy quark velocity). Thus the terms which are proportional to $(\pi/v)$ or $(\pi/v)^2$ should be treated separately in that different PMC scales are adopted in the estimation~\cite{brodsky1,pmc3}.

\item The factorization scale $\mu_f$ which enters into the predictions for QCD inclusive reactions is introduced to match nonperturbative and perturbative aspects of the parton distributions in hadrons. The factorization scale occurs even for a conformal theory with $\{\beta^{\cal R}_i\}=0$ where $\alpha_s$ is constant. The factorization scale should be chosen to match the nonperturbative bound state dynamics with perturbative DGLAP evolution. This can be done explicitly for electron-atom or atom-atom inelastic scattering processes in QED using the known bound state dynamics of atoms. This could also be done in hadron physics using nonperturbative models such as AdS/QCD and light-front holography; recent reviews can be found in Refs.~\cite{ads1,ads2}. There is clearly no reason to equate the factorization scale to the renormalization scale~\cite{stanwu}. We expect that the factorization scale ambiguity will also be reduced by applying the PMC scale-setting to the kernels of DGLAP evolution equations.

\end{itemize}

In summary, the systematic application of the PMC can eliminate a major ambiguity of pQCD predictions from scale and scheme ambiguities, thus greatly improving the precision of tests of the Standard Model and the sensitivity to new physics at  colliders and other experiments.

\begin{acknowledgments}
We thank Hung-Jung Lu, Leonardo di Giustino, George Sterman, Andrei L. Kataev, Sergey V. Mikhailov, Zvi Bern, Stefan Hoeche, and Dmitry V. Shirkov for helpful conversations. This work was supported in part by the Program for New Century Excellent Talents in University under Grant NO.NCET-10-0882, Natural Science Foundation of China under Grant NO.11075225 and NO.11275280
, and the Department of Energy contract DE-AC02-76SF00515. SLAC-PUB-15133.
\end{acknowledgments}

\end{document}